\documentclass[aps,prd,twocolumn,nofootinbib,preprintnumbers,floatfix,superscriptaddress,notitlepage]{revtex4-1}

\usepackage{graphicx}
\usepackage{amsmath,amsfonts,amssymb}
\usepackage[breaklinks,colorlinks,urlcolor=black,citecolor=black,linkcolor=black]{hyperref}
\usepackage{color}
\usepackage{verbatim}

\def\lsim{\mathrel{\raise.3ex\hbox{$<$\kern-.75em\lower1ex\hbox{$\sim$}}}}
\def\gsim{\mathrel{\raise.3ex\hbox{$>$\kern-.75em\lower1ex\hbox{$\sim$}}}}

\newcommand{\be}{\begin{equation}}
\newcommand{\ee}{\end{equation}}
\newcommand{\bea}{\begin{equation}\begin{aligned}}
\newcommand{\eea}{\end{aligned}\end{equation}}
\newcommand{\td}{{\rm d}}
\newcommand{\eg}{{e.g.}}
\newcommand{\ie}{{i.e.}}
\newcommand{\Msun}{M_{\odot}}

\newcommand{\kpc}{{\rm kpc}}

\newcommand{\Gpc}{{\rm Gpc}}
\newcommand{\yr}{{\rm yr}}

\begin{document}

\title{Prospects for probing gravitational waves from primordial black hole binaries}

\author{Oriol Pujolas}
\email{pujolas@ifae.es}
\affiliation{Institut de Fisica d'Altes Energies (IFAE), The Barcelona Institute of Science and Technology, Campus UAB, 08193 Bellaterra (Barcelona), Spain}

\author{Ville Vaskonen}
\email{vvaskonen@ifae.es}
\affiliation{Institut de Fisica d'Altes Energies (IFAE), The Barcelona Institute of Science and Technology, Campus UAB, 08193 Bellaterra (Barcelona), Spain}

\author{Hardi Veerm\"ae}
\email{hardi.veermae@cern.ch}
\affiliation{National Institute of Chemical Physics and Biophysics, R\"avala 10, 10143 Tallinn, Estonia}

\begin{abstract}
We study the prospects of future gravitational wave (GW) detectors in probing primordial black hole (PBH) binaries. We show that across a broad mass range from $10^{-5}M_\odot$ to $10^7M_\odot$, future GW interferometers provide a potential probe of the PBH abundance that is more sensitive than any currently existing experiment. In particular, we find that galactic PBH binaries with masses as low as $10^{-5}M_\odot$ may be probed with ET, AEDGE and LISA by searching for nearly monochromatic continuous GW signals. Such searches could independently test the PBH interpretation of the ultrashort microlensing events observed by OGLE. We also consider the possibility of observing GWs from asteroid mass PBH binaries through graviton-photon conversion.
\end{abstract}

\maketitle

\section{Introduction}

The advent of gravitational wave (GW) astronomy~\cite{Abbott:2016blz} has revolutionized astrophysics and cosmology. One consequence of the LIGO-Virgo detections has been the re-emergence of interest in primordial black holes (PBHs). In particular, various groups have studied the merger rate of PBHs and the consequences of the LIGO-Virgo detections~\cite{LIGOScientific:2018mvr,LIGOScientific:2020ibl} on PBHs, by considering the possibility that at least some of the observed events include PBHs or setting stringent constraints on their abundance~\cite{Sasaki:2016jop,Bird:2016dcv,Clesse:2016vqa,Raidal:2017mfl,Ali-Haimoud:2017rtz,Raidal:2018bbj,Vaskonen:2019jpv,Gow:2019pok,Young:2019gfc,DeLuca:2020jug,Jedamzik:2020ypm,Hall:2020daa,Young:2020scc,Wong:2020yig,Kritos:2020wcl,DeLuca:2020qqa,Hutsi:2020sol,DeLuca:2021wjr,Franciolini:2021tla}. Currently the consensus is that the scenario where all observed events are of primordial origin is disfavoured by the LIGO-Virgo data, and the abundance of $\mathcal{O}(10\Msun)$ PBHs is bounded to be smaller than $\mathcal{O}(0.1\%)$ of the total dark matter (DM) abundance~\cite{Hutsi:2020sol}. However, having a sub-population of PBH binaries is consistent with latest observations~\cite{Hutsi:2020sol,DeLuca:2021wjr,Franciolini:2021tla}.

Whereas the present LIGO-Virgo detectors are sensitive to mergers of $\mathcal{O}(0.1-100)\Msun$ compact objects, various detectors are currently being designed that allow us to probe a wide mass range from well below the astrophysical BH mass bound to extremely heavy BHs seen in galactic centers~\cite{Punturo:2010zz,LISA:2017pwj,Graham:2017pmn,Badurina:2019hst,Bertoldi:2019tck}. For example, these detectors will provide invaluable information about the mass, spin, and redshift distributions of the BH merger rate, which will help us determine the origin of these BHs. Particularly interesting are the prospects of probing subsolar mass BHs.\footnote{A search for sub-solar mass PBH binaries using the Ligo-Virgo O2 data found four candidate events although they were not sufficiently statistically significant to count them as detections~\cite{Phukon:2021cus}.} Such BHs are not expected to be formed through usual stellar evolution, and therefore their detection would be a strong signature of a PBH population.\footnote{Non-primordial subsolar mass PBHs could form in cosmologies with a dissipative dark sector~\cite{Chang:2018bgx,Shandera:2018xkn} or via DM induced collapse of neutron stars or white dwarfs~\cite{Kouvaris:2018wnh,Dasgupta:2020mqg}.}

The abundance of PBHs is constrained by several different observations (for a review, see \eg~\cite{Carr:2020gox}). The abundance of the lightest non-evaporated PBHs is bounded by the non-observation of radiation originating from their slow evaporation~\cite{Carr:2009jm,Raidal:2018eoo,Laha:2020ivk,Coogan:2020tuf,Ray:2021mxu,Mittal:2021egv}. For higher masses, the non-observation of microlensing events implies strong constraints on the PBH abundance~\cite{Tisserand:2006zx,Niikura:2017zjd,Smyth:2019whb,Niikura:2019kqi}. There is, however, a window between the evaporation and the microlensing constraints, roughly in the asteroid mass range from $10^{16}$\,g to $10^{23}$\,g where PBHs may comprise all DM. In light of the current constraints, this is also the only mass window where all DM could be in PBHs: For higher masses lensing of type Ia supernovae~\cite{Zumalacarregui:2017qqd}, the LIGO-Virgo GW observations~\cite{Raidal:2017mfl,Ali-Haimoud:2017rtz,Raidal:2018bbj,Vaskonen:2019jpv,DeLuca:2020qqa,Hutsi:2020sol}, PBH accretion~\cite{Ricotti:2007au,Ali-Haimoud:2016mbv,Poulin:2017bwe,Hektor:2018qqw,Serpico:2020ehh}, survival of a stars in dwarf galaxies~\cite{Brandt:2016aco,Koushiappas:2017chw}, and Lyman-$\alpha$ forest data~\cite{Afshordi:2003zb,Murgia:2019duy} imply strong constraints on the PBH abundance.

Another interesting subsolar PBH mass range is around $10^{-5}\Msun$, where the OGLE experiment has found six ultrashort microlensing events~\cite{2017Natur.548..183M}. Combined with the microlensing constraints from Subaru-HSC~\cite{Niikura:2017zjd}, these events can be explained if $\mathcal{O}(10^{-5}\Msun)$ PBHs comprise $\mathcal{O}(1\%)$ of the total DM abundance~\cite{Niikura:2019kqi}.

In this paper, we study the prospects of future GW detectors to probe PBHs, paying particular attention to subsolar mass PBHs. We consider GW signals from individual PBH binaries as well as the stochastic GW background (SGWB) generated by those PBH binaries that are not individually resolvable. We find that searches of nearly monochromatic continuous GWs from PBH binaries within the Milky Way DM halo greatly improves the prospects for probing the abundance of subsolar mass PBHs. Finally, we consider observing asteroid mass PBHs using graviton-to-photon conversion but find that the induced signal is too weak to be detectable with near-future technology.

\section{PBH merger rate}

PBHs form binaries very efficiently in the early universe after they decouple from the Hubble flow~\cite{Nakamura:1997sm,Ioka:1998nz,Raidal:2017mfl,Ali-Haimoud:2017rtz,Raidal:2018bbj,Vaskonen:2019jpv}. At time $t$, the merger rate of PBH binaries is
\be \label{eq:R}
	R(t) \!=\! \frac{3.0\times 10^6}{\Gpc^{3}\yr} f_{\rm PBH}^{\frac{53}{37}} \!\left[\frac{m}{M_\odot}\right]^{\!-\frac{32}{37}} \!\left[\frac{t_0}{t}\right]^{\!\frac{34}{37}} \!S(f_{\rm PBH})\,, 
\ee
where $t_0$ denotes the age of the Universe, $m$ is the PBH mass\footnote{For simplicity, we assume a monochromatic PBH mass function. A more general expression for the merger rate is reviewed e.g. in Ref.~\cite{Hutsi:2020sol}.}, $f_{\rm PBH}$ is the fraction of DM in PBHs, and $S<1$ is a suppression factor. The latter can be divided into two parts, $S = S_1 S_2$: The first one excludes initial configurations where a third PBH likely falls into the binary and includes the effect of perturbations in the ambient smooth matter component. $S_1$ is independent of $t$, but depends non-trivially on the mass function, the variance of matter density fluctuations and the PBH abundance~\cite{Raidal:2018bbj}. We will use the approximate $S_1$ derived in Ref.~\cite{Hutsi:2020sol}. For $10^{-4}\leq f_{\rm PBH}\leq1$, this factor is in the range $0.16 < S_1 < 0.47 $ and for smaller abundances it decreases as $f_{\rm PBH}^{21/37}$ so that then $R \propto f_{\rm PBH}^{2}$. The second factor, $S_2$, excludes binaries that become part of a DM halo in which close encounters with other PBH are very likely. This suppression factor accounts for the enhanced small-scale structure formation in PBH cosmologies~\cite{Carr:2018rid,Hutsi:2019hlw, Inman:2019wvr} and depends on the time when the binary merges. At $t=t_0$ it can be approximated by~\cite{Vaskonen:2019jpv}
\be \label{eq:S2}
    S_2(t\!=\!t_0) = \min\left[1,\,9.6\times 10^{-3} f_{\rm PBH}^{-0.65} e^{0.03 \ln^2\!f_{\rm PBH}}\right]\,.
\ee
For binaries merging at time $t$, the factor $S_2$ is obtained by replacing $f_{\rm PBH}\to f_{\rm PBH} (t/t_0)^{0.44}$ in~\eqref{eq:S2}. For $f_{\rm PBH}=1$ the present ($t=t_0$) total suppression factor is $S = 2.3\times 10^{-3}$.

We assume that PBHs are not initially clustered. Initial clustering generally enhances formation of the initial PBH binaries~\cite{Raidal:2017mfl,Ballesteros:2018swv,Young:2019gfc,DeLuca:2021hde}, but it can also increase the probability for these binaries to be disrupted later in DM haloes~\cite{Raidal:2018bbj,Vaskonen:2019jpv,Jedamzik:2020ypm,Jedamzik:2020omx}. The latter thus suppresses the merger rate of PBH binaries from this formation channel. At the same time, initial clustering can boost the merger rate of other PBH binary formation channels, which are subdominant in the initially Poissonian case, \eg, of disrupted initial PBH binaries~\cite{Vaskonen:2019jpv} or of binaries formed later in DM haloes~\cite{Clesse:2016vqa,Raidal:2017mfl,Jedamzik:2020ypm,Jedamzik:2020omx}. With strong initial clustering, the future GW interferometers may probe PBH abundances as low as $f_{\rm PBH} = 10^{-10}$~\cite{DeLuca:2021hde}.

\begin{figure}
\centering
\includegraphics[height=0.224\textwidth]{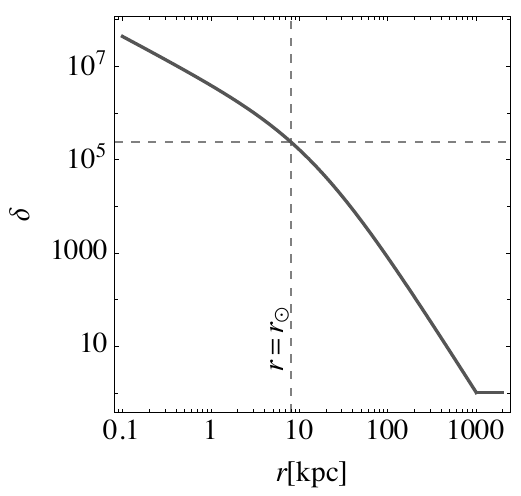}
\includegraphics[height=0.224\textwidth]{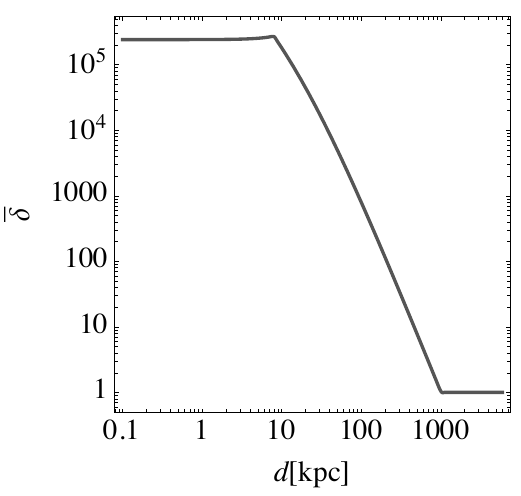}
\caption{\emph{Left panel:} The density contrast of the Milky Way DM halo as a function of radius. \emph{Right panel:} The average DM density contrast as a function of the distance from the Solar System.}
\label{fig:Rdelta}
\end{figure}

For subsolar mass PBH binaries, the overdensity of the Milky Way DM halo has to be taken into account. Assuming that the present density of PBH binaires follows the density of the rest of the DM, this simply gives an extra factor of $\delta(r) \equiv \rho_{\rm DM}(r)/\bar\rho_{\rm DM}$ for the merger rate. We model the Milky Way DM halo by the Navarro-Frenk-White density profile~\cite{Navarro:1995iw,Navarro:1996gj},
\be
	\rho_{\rm DM}(r) = \frac{\rho_0}{\frac{r}{r_0} \left(1+\frac{r}{r_0}\right)^{\!2}} \,,
\ee
with $r_0 = 15.6\kpc$. The reference energy density $\rho_0$ is chosen such that the local ($r_\odot = 8.0\,\kpc$) DM density is $\rho_{\rm DM}(r\!=\!r_\odot) = 7.9\times10^{-3}\Msun/{\rm pc}^3$~\cite{Cautun:2019eaf}. This density profile is shown in the left panel of Fig.~\ref{fig:Rdelta}. At large distances where $\rho_{\rm DM}(r) < \bar\rho_{\rm DM}$ we take $\delta(r) = 1$. The average DM density contrast at distance $d$ from the Solar System is
\be 
	\bar\delta(d) = \max\left[1, \,\frac{1}{2} \int_{-1}^1 \!\td\!\cos\theta \,\delta(r)\right] \,,
\ee
where $r^2 = d^2 +r_\odot^2 - 2\cos\theta\, d\, r_\odot$. We show $\bar\delta(d)$ in the right panel of Fig.~\ref{fig:Rdelta}. For example, assuming $f_{\rm PBH}=1$ the total number of PBHs within the virial radius ($\delta=200$) in the Milky Way halo is $1.4\times 10^{25}\, [10^{20}\,{\rm g}/m]$ and their total present ($t=t_0$) present merger rate is $3/{\rm yr}\, [10^{-5}\Msun/m]^{32/37}$. Note that, since light nearby binaries can be observable for a much longer period than a year (see Fig.~\ref{fig:d8}), then the number of binaries that are observable at each instant can be much larger.

\begin{figure*}
\centering
\includegraphics[width=0.48\textwidth]{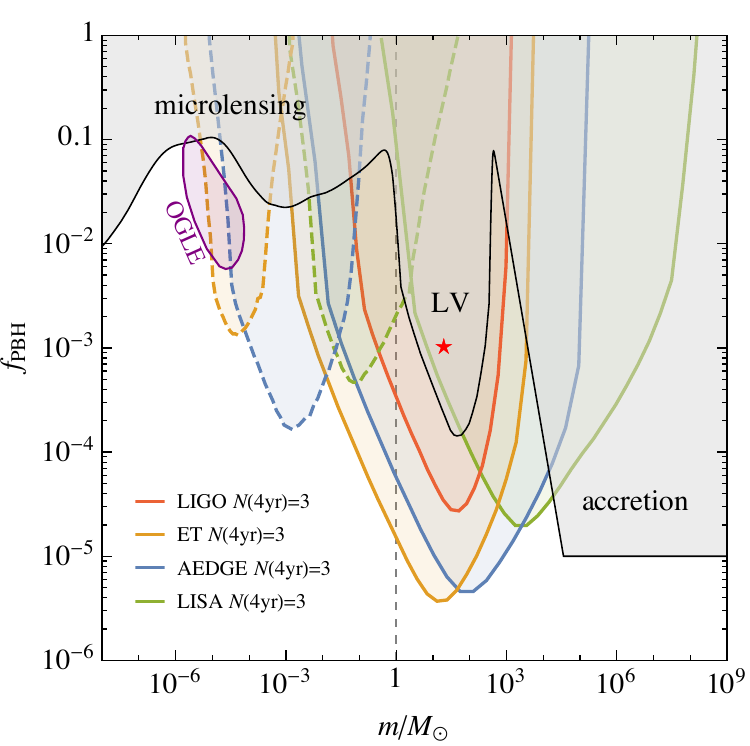} 
\includegraphics[width=0.48\textwidth]{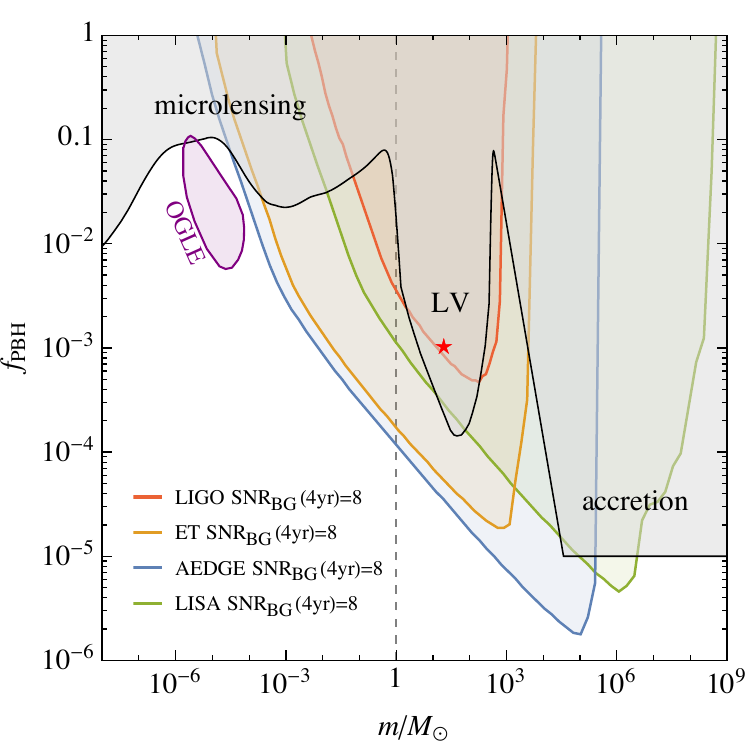}
\caption{Sensitivities of future GW observatories on the PBH abundance through GWs from PBH binaries. The left panel corresponds to observations of GWs from individual PBH binaries, and the right panel to the SGWB from PBH binaries. The thin black line shows the envelope of the current PBH constraints, and the red star the scenario in which all LIGO-Virgo BHs are primordial. The purple region corresponds to the PBH explanation of the OGLE excess events.}
\label{fig:sensproj}
\end{figure*}

\section{GW interferometers}

Three qualitatively different types of GW signals can originate from BH binaries: individual mergers appearing in relatively brief transient signals, inspiralling binaries causing a nearly monochromatic signal that can last much longer than the observational period, and individually unresolvable binaries contributing to the SGWB. We will consider each case separately and study how the PBH hypothesis can be tested by searching for such GW signals with future GW interferometers.

\subsection{Short-duration transient signals}

Let us first consider relatively short GW signals from individual PBH binary mergers. The expected number of PBH binary mergers that a GW detector, whose sensitivity is characterized by a noise power spectrum $S_n(f)$, should see within the period $T$ is
\bea
	N = T \int \frac{\td z}{1+z}\, \frac{\td V_c}{\td z}& \,\bar\delta(d(z)) R(t(z)) \\ &\times p_{\rm det}({\rm SNR}_c/{\rm SNR}(z)) \,,
\eea  
where $V_c(z)$ is the volume of the comoving Hubble horizon at redshift $z$ and $p_{\rm det}$ is the detection probability, which accounts for the antenna patterns of the detector and averages over the binary inclination and sky location, and the signal polarization~\cite{Finn:1992xs,Gerosa:2019dbe}. The detectability of a signal $\tilde h$ is characterized by the signal-to-noise ratio
\be \label{eq:SNR1}
    {\rm SNR}(z) = \sqrt{\int_0^\infty \td f\, \frac{4|\tilde{h}(f)|^2}{S_n(f)}} \,.
\ee 
We use ${\rm SNR}_c=8$ as the threshold value for detection. For the Fourier transform of the signal, $\tilde{h}(f)$, we use the approximation given in Ref.~\cite{Ajith:2007kx}. Using $p_{\rm det}$, the signal-to-noise ratio is evaluated for the optimally oriented source.

The projected sensitivities of the LIGO~\cite{LIGOScientific:2014pky}, ET~\cite{Hild:2010id} and LISA~\cite{Cornish:2018dyw} to GW bursts from PBH mergers are depicted in Fig.~\ref{fig:sensproj} by the solid colored curves. Above each solid curve, we expect to see at least 4 PBH binary signals within an observation period of 4 years, $N(T = 4\,{\rm yr}) >3$. If no PBH binary signals are observed, these correspond to 95\% confidence level upper limits on the PBH abundance. We see that these experiments will be sensitive to very small PBH abundances, even down to $f_{\rm PBH}\sim 10^{-5}$, in the mass range from $10^{-3}M_\odot$ to $10^8M_\odot$.

The black solid curve in Fig.~\ref{fig:sensproj} shows the envelope of the current PBH constraints. It includes the constraints arising from microlensing results from Subaru-HSC~\cite{Niikura:2017zjd,Smyth:2019whb}, EROS~\cite{Tisserand:2006zx} and OGLE~\cite{2017Natur.548..183M,Niikura:2019kqi}, LIGO-Virgo observations~\cite{Hutsi:2020sol}, and  accretion~\cite{Ali-Haimoud:2016mbv,Serpico:2020ehh}. For the LIGO-Virgo constraints, we have neglected the observed BH binary merger events. The current observations are consistent with a subpopulation of the observed events originating from PBH mergers~\cite{Hutsi:2020sol,Hall:2020daa,DeLuca:2021wjr,Franciolini:2021tla}. However, to consistently include the possibility that some of these events had a primordial origin, we must account for the shape of the mass function.\footnote{Since the merger rate is not linear in the mass function, there is no simple prescription to generalize the monochromatic estimates to general mass functions~\cite{Carr:2017jsz}. Nevertheless, the method described in~\cite{Carr:2017jsz} works well when the mass function is sufficiently narrow.} The monochromatic mass function assumed here is not consistent with observations. The most recent PBH constraints accounting for the possibility that some of the observed PBH may be primordial are given in~\cite{Hutsi:2020sol}. Compared to the case where all observed mergers are astrophysical, these constraints are softened by less than an order of magnitude in the $2-100 \Msun$ mass range -- roughly, the LIGO-Virgo constraints are removed below the abundance at the red star in Fig.~\ref{fig:sensproj}.

\begin{figure*}
\centering
\includegraphics[width=0.98\textwidth]{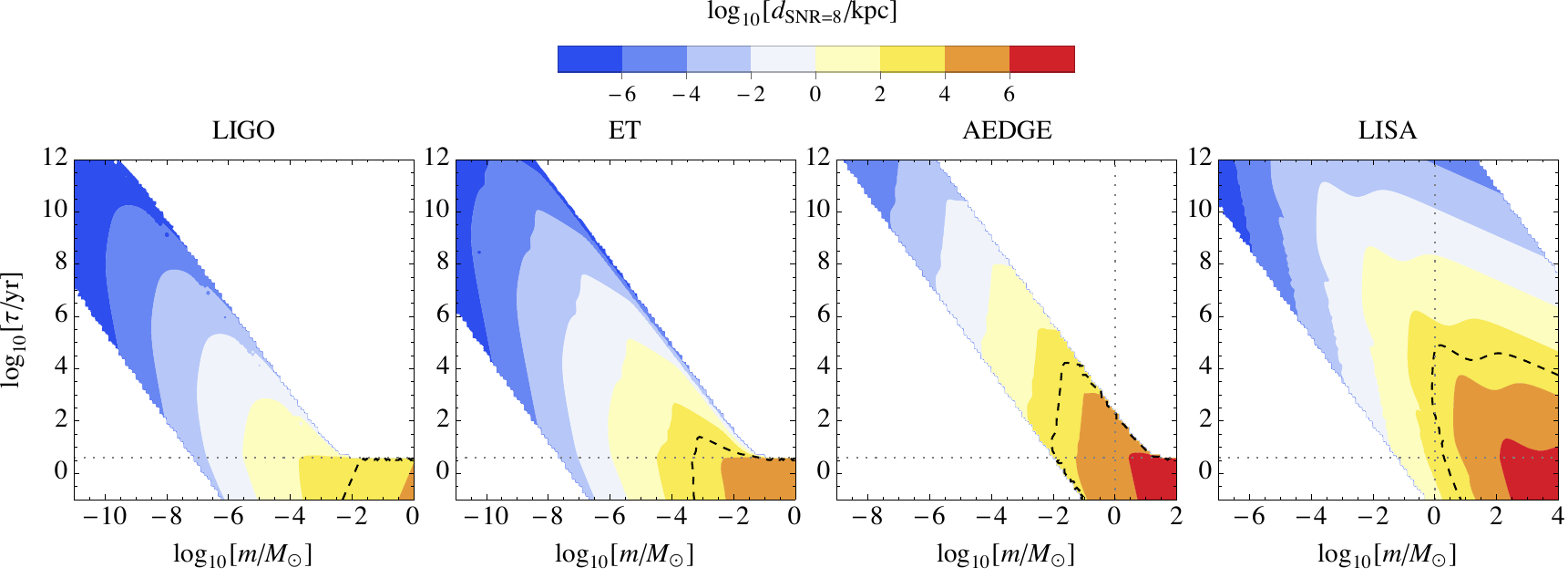} 
\caption{The color coding indicates the maximal distance at which ${\rm SNR}>8$ from a circular equal mass BH binary whose coalescence time is $\tau$. The dashed contour indicates the projection of the boundary of the Milky Way DM halo. In the white regions, the experiment is not sensitive to the signal frequency.}
\label{fig:d8}
\end{figure*}

\subsection{Nearly monochromatic continuous signals}

Nearly monochromatic continuous signals from isolated spinning neutron stars have been searched, but not found, in the LIGO-Virgo data~\cite{Piccinni:2019zub,LIGOScientific:2019yhl,LIGOScientific:2021mwx,LIGOScientific:2021tsm}. Similar signals can arise from light binaries for which the merger frequency is too large, so only the inspiral phase is observable~\cite{Horowitz:2019pru,Miller:2020kmv}. In this case, we assume that the signal lasts for the entire duration $T$ of the experiment, \ie, the coalescence time $\tau$ satisfies $\tau > T$, and we count the potential PBH binaries that emit GWs in the frequency range accessible to the detector.

The present comoving number density at $t_0$ of binaries that will merge after a time $\tau$ is $\bar\delta(d) R(t_0 + \tau) \td \tau$. Assuming that the orbits are nearly circularized by GW emission, a binary with coalescence time $\tau$ will emit GWs at the frequency
\be
    f(\tau) = \frac{\left(5/\tau\right)^{\frac38}}{8\pi \left(G \mathcal{M}\right)^{5/8}} \,,
\ee
where $\mathcal{M}$ denotes the chirp mass which for equal mass binary is $\mathcal{M} = 2^{-1/5} m$. The number of detectable local sources emitting at frequency $f$ is
\bea\label{eq:dNdf}
    \frac{\td N}{\td f} 
    = \int \td V_c \td \tau \, \delta(f& - f(\tau)) \bar\delta(d) R(t_0 + \tau) \\
    & \times p_{\rm det}({\rm SNR}_c/{\rm SNR}(d,\tau)) \,,
\eea
where the volume integral is taken over the galactic sources only, so we can neglect cosmic expansion. As with short signals, we use $p_{\rm det}$ to estimate the detectability. We note that, for long-lasting signals, the movement and reorientation of the detector with respect to the source causes daily modulation in the observed signal that may help in separating it from the detector noise. 

Due to the computational cost of fully coherent all-sky searches of continuous GWs, various semi-coherent methods have been developed (see \eg~\cite{Jaranowski:1998qm,Krishnan:2004sv,Prix:2012yu,Astone:2014esa}). For simplicity, assuming that well optimised semi-coherent methods can reach nearly similar results, we calculate the most optimal signal-to-noise ratio\footnote{The apparent factor of $\sqrt{2}$ difference compared to Eq.~\eqref{eq:SNR1} arises as Eq.~\eqref{eq:SNR1} is written such that the frequency integral is from $0$ to $\infty$.}
\be
    {\rm SNR}(d,\tau) = \sqrt{\int_{0}^{T} \!\!\td t\, \frac{2|h(t)|^2}{S_n(f(\tau-t))}} \,,
\ee
where
\be
    |h(t)| = \frac{4}{\pi^{\frac23} d} \left(G \mathcal{M}\right)^{\frac53} f(\tau-t)^{\frac23} 
\ee
is the amplitude of the GW signal from a circular BH binary inspiral. 

The expected sensitivities of future GW observatories, in terms of the distance $d$ up to which the signal can be seen with ${\rm SNR}>8$, are shown in Fig.~\ref{fig:d8}. Since the observable binaries, which by Fig.~\ref{fig:sensproj} are heavier than $10^{-6}\Msun$, are expected to merge much faster than a Hubble time, $\tau \ll t_0$, we can approximate $R(t_0 + \tau) \approx R(t_0)$. The shape of the frequency distribution of sources is thus practically flat when $\tau \ll t_0$ and the observed distribution of sources~\eqref{eq:dNdf} will follow the sensitivity of the experiment. However, any future experiments that would be sensitive to local binaries expected to coalesce in a time longer than a Hubble time might probe the shape of the coalescence time distribution, which follows an universal power law $R \propto (t+\tau)^{34/37}$ for PBH mergers.

The total expected number of detectable sources is
\bea
    N  = 4\pi R(t_0) \int \td d \, d^2\,& \td \tau \,  \bar\delta(d)  \\
    & \times p_{\rm det}({\rm SNR}_c/{\rm SNR}(d,\tau)) \,.
\eea
The colored dashed contours in the left panel of Fig.~\ref{fig:sensproj} show the projected sensitivities of the future GW experiments to nearly monochromatic signals from PBH binaries. The effect of the Milky Way DM halo on the expected number of events is relevant for these observations and allows for ET, AEDGE and LISA to probe subsolar mass PBH binaries. We find that the design sensitivity of LIGO is not sufficient for probing a potential galactic PBH binary population. Our result agree with the semi-coherent analysis in Ref.~\cite{Miller:2020kmv} that considered searches of continuous signals from light PBHs in the Milky Way with LIGO and ET. Compared to Ref.~\cite{Miller:2020kmv}, the tails of the dashed ET sensitivity curve in Fig.~\ref{fig:sensproj} are slightly steeper. This difference arises mainly because we use a more conservative suppression factor in the PBH merger rate~\eqref{eq:R}.

OGLE reported six ultrashort-timescale microlensing events~\cite{2017Natur.548..183M} that may have originated from PBHs~\cite{Niikura:2019kqi}. The corresponding PBH mass-abundance region is shown in purple in Fig.~\ref{fig:sensproj}. From the left panel, we see that the PBH explanation for these events may be tested by searches of continuous signals from galactic PBH binaries with ET or AEDGE.

Finally, let us take a closer look at the distribution of orbital eccentricities. This may be used as a potential tool for testing the origin of the binaries, and is especially relevant for BHs above a few solar masses for which astrophysical formation channels exist. PBH binaries formed in the early Universe have initially highly eccentric orbits. The initial distribution of the dimensionless angular momentum $j \equiv \sqrt{1-e^2}$, where $e$ is the eccentricity, for binaries with initial coalescence time $t_0+\tau$ is approximately~\cite{Raidal:2018bbj}
\be
    P_{\rm init}(j|t_0+\tau) 
    \propto j^{\frac{56}{16}} e^{-j^{\frac{37}{8}}/\alpha_0} \,,
\ee
where  
\bea
    \alpha_0
    = 4 \times& 10^{-9}\, 
    \left[\frac{M}{\Msun}\right]^{\frac{5}{8}} \left[1+\frac{\tau}{t_0}\right]^{\frac{3}{8}} \\
    &\times 
    \left( f_{\rm PBH}^{2}+ f_{\rm PBH} \sigma_M + 2\sigma_M^2\right) \,.
\eea
The variance of matter density fluctuations at the time of PBH binary formation is $\sigma_M \approx 0.005$. This approximation works best when $f_{\rm PBH} \gtrsim \sigma_M$ which is roughly satisfied for observable abundances. For $\tau \ll t_0$, the $j$ distribution is practically independent of $\tau$. For initially highly eccentric orbits, $j$ evolves approximately as $j = j_0 (1 + t_0/\tau)$~\cite{PhysRev.136.B1224}, which, when $\tau \ll t_0$, corresponds to
\be
    \alpha_t \approx \alpha_0 \left(t_0/\tau\right)^{\frac{37}{8}} \,.
\ee
The time at which the binaries have circularized ($j\approx1$) can be estimated from $\alpha_t \approx 1$.\footnote{These approximations cannot describe the distribution of nearly circular present-day binaries. In particular, since $j\leq1$ by construction, the distribution is unphysical for $\alpha \gtrsim 1$, and one must more carefully evaluate the binary evolution due to GW emission.} For example, when $M = 1\Msun$ and $f_{\rm PBH} =1$, eccentric binaries must have a present-day coalescence time $\tau \gtrsim 200$\,Myr. Using Fig.~\ref{fig:d8} we can then conclude that $M > 1\Msun$ binaries accessible by LIGO, ET and AEDGE will have circularized due to GW emission.

\subsection{Stochastic background}

Finally, we consider all those binaries that a given detector can not individually resolve, but contribute to a SGWB. The SGWB from PBH binaries in light of the LIGO-Virgo data has been considered in several works~\cite{Wang:2016ana,Mandic:2016lcn,Cholis:2016xvo,Raidal:2017mfl,Raidal:2018bbj,Vaskonen:2019jpv,Wang:2019kaf,Hutsi:2020sol,DeLuca:2020qqa,Mukherjee:2021ags}. The strength of this SGWB is given by
\be \label{eq:GWB}
	\Omega_{\rm GW}(f) = \int \frac{\td z}{1+z}  \frac{\td V_c}{\td z} \,\bar\delta(d(z)) R(t(z))  \frac{1}{\rho_c} \frac{\td \rho_{\rm GW}}{\td f} 
	\,,
\ee
where 
\be
    \td \rho_{\rm GW} = \frac{4}{25}\frac{\pi}{G} \,f^3 |\tilde{h}(f)|^2 \td f
\ee
is the GW energy density emitted by a binary in the frequency range $(f,f+\td f)$~\cite{Moore:2014lga}, and $\rho_c$ denotes the critical density. We estimate the detectability of the SGWB from the signal-to-noise ratio
\be
    {\rm SNR}_{\rm BG} = \sqrt{ T \int \td f \left[\frac{\Omega_{\rm GW}(f)}{\Omega_n(f)}\right]^2} \,,
\ee
where $\Omega_n(f) = \pi f^3 S_n(f)/(4\rho_c)$ is the dimensionless energy density in noise~\cite{Moore:2014lga}. We assume that signals with ${\rm SNR}_{\rm BG} > {\rm SNR}_c$ are detectable.

In the right panel of Fig.~\ref{fig:sensproj} we show the regions where different GW interferometers can detect the SGWB from PBH binaries. By comparing with the left panel, where the sensitivities for GW signals from individual PBH binaries are shown, we see that also the SGWB provides a powerful probe of subsolar mass PBHs, in agreement with~\cite{Mukherjee:2021itf,DeLuca:2021hde}. However, it will be difficult to determine the masses of the binaries if, in the frequency range of these detectors, only the inspiral part of the spectrum contributes to the SGWB. The SGWB will provide the most sensitive probe of the PBH abundance at high masses $10^3M_\odot < m < 10^7M_\odot$.

\section{Graviton-to-photon conversion}

GWs can be converted into photons in a transverse magnetic field through the inverse Gertsenshtein process~\cite{Gertsenshtein:1961}. This offers an alternative channel for observing GWs in frequency ranges not accessible to GW interferometers and has thus been mostly considered for GWs with frequencies above GHz~\cite{Cruise:2012zz,Ejlli:2019bqj,Ringwald:2020ist}. Such high-frequency signals can be generated by light PBH binaries~\cite{Herman:2020wao}. Asteroid mass PBHs are particularly interesting as their abundance is not constrained, and thus they could make up all of DM~\cite{Carr:2020gox}. So far, this range may be tested only indirectly by observing the SGWB associated with the formation of such light PBHs (see \eg~\cite{Saito:2008jc,Assadullahi:2009jc,Bugaev:2010bb,Alabidi:2012ex,Orlofsky:2016vbd,Espinosa:2018eve,Inomata:2018epa,Byrnes:2018txb,Clesse:2018ogk,Cai:2018dig,Wang:2019kaf,Chen:2019xse}). This SGWB is, however, strongly tied to the details of the PBH formation scenario and thus its (non-)observation cannot be conclusively tied to the (non-)existence of PBHs.

As an idealized GW detector, take a cylinder of length $L$ and cross section $A$, with a magnetic field of strength $\bar B_x$ perpendicular to the symmetry axis of the cylinder. For $2\pi f L \gg 1$ the photon power through the area $A$ induced by a GW traveling perpendicular to the magnetic field is~\cite{Boccaletti1970,osti_4377804,DeLogi:1977qe}
\be \label{eq:Ps}
    P_s(f) \approx \frac{\pi^2}{4\mu_0 c} \bar B_x^2 A L^2 f^2 |h_c(f)|^2 \,,
\ee
where $|h_c(f)| = 2 f |\tilde h(f)|$ is the dimensionless characteristic strain of the GW. The photon power may be resonantly enhanced in narrow frequency bands, determined by the length of the detector, by up to 6 orders of magnitude~\cite{Ejlli:2019bqj}. However, in the search of transient GW signals the resonant enhancement is useful only if the frequency of the signal matches one of the resonant frequencies and changes very slowly.

Consider now the GW signal from an asteroid mass PBH binary within the Milky Way halo. The GW frequency emitted by a merger of an equal mass BH binary is very high $f_{\rm merger}\!\approx\! 1.6\times 10^{17}\,{\rm Hz} \,[10^{20}\,{\rm g}/m]$, and it is therefore reasonable consider only the inspiral signal, 
\be \label{eq:hc}
	|h_c(f)| = \sqrt{\frac{5}{6}} \frac{\left(G\mathcal{M}\right)^{\frac56}}{\pi^{\frac23} d} f^{-\frac16} \,.
\ee 
Combining \eqref{eq:Ps} and \eqref{eq:hc}, such a signal produces photons at the rate
\bea
    \Gamma_\gamma(t) \equiv \frac{P_s(f)}{2\pi \hbar f} = &\frac{0.02}{s} \!\left[\frac{\bar B_x}{10\,{\rm T}}\right]^2 \!\left[\frac{L}{\rm m}\right]^2  \!\left[\frac{A}{{\rm m}^2}\right] \\ 
    &\times\!\left[\frac{m}{10^{20}\,{\rm g}}\right]^{\!\frac53} \!\left[\frac{f}{{\rm GHz}}\right]^{\!\frac23} \!\left[\frac{d}{{\rm 50 au}}\right]^{\!-2}\,
\eea
in the detector described above. The coalescence time of a binary that presently emits GWs at frequency $f$ is
\be
	\tau(f) \approx 0.14\,{\rm day}\, \left[\frac{m}{10^{20}{\rm g}}\right]^{-\frac53} \left[\frac{f}{\rm GHz}\right]^{-\frac83} \,.
\ee
Thus, although increasing frequency increases the number of photons, it will shorten the time window for observing the signal.

As an extreme example, assume a PBH binary with $m=10^{20}$\,g in the Kuiper belt, $d=50$\,au, and a detector with $B=10$\,T, $L = 1$\,m, and $A=1\,{\rm m}^2$ that can probe frequencies from $0.3$\,GHz to $3$\,GHz. The GW signal from the inspiral will stay in the sensitivity window for $3.4$\,days\footnote{This binary would be microscopic, with a $\mathcal{O}(10^{2}\mu \rm m)$ semimajor axis.}, and induces in total thousands of photons in the detector. Assuming a background event rate $\mathcal{O}(1\, \rm mHz)$, comparable to current and proposed detectors~\cite{Ejlli:2019bqj}\footnote{These detectors are sensitive to much higher frequencies.}, this signal would be detectable. The existence of such accidental nearby sources is, however, unlikely, and the probability of observation can be further diminished due to the brief time window when the signal is in the sensitivity range of the detector.

To estimate the number of PBH binaries accessible to a detector described above, we will, for simplicity, assume that $f_{\rm PBH}=1$. The expected number of PBH binaries emitting at a frequency $f$ that are observable during a time interval $T$ and within a distance $d\ll r_\odot$ from us is approximately
\bea \label{eq:Napr}
	N(d) &\approx \frac{4\pi}{3} d^3\,T\,\delta(r_\odot) R(t_0 + \tau(f)) \\
	&\approx 2.2\times 10^3\,  \left[\frac{T}{\rm yr}\right] \left[\frac{d}{\rm kpc}\right]^3 \left[\frac{m}{10^{20}\,{\rm g}}\right]^{-\frac{32}{37}} \,,
\eea
where $\delta(r_\odot)$ is the local DM density contrast. Eq.~\eqref{eq:Ps} assumes that the direction of the GW is perpendicular to the magnetic field inside the detector. To calculate the expected number of observed events, we must account for the limited sensitivity of the detector on GWs that approach from other directions. This can be well estimated by considering only GWs approaching the detector from the solid angle $\tilde \Omega = A/(L/2)^2$ above the cylinder. The expected number of observed events is therefore
\be
    \tilde N(d) \equiv \frac{\tilde \Omega}{4\pi} \,N(d) = \frac{A N(d)}{\pi L^2} \,.
\ee 
To see at least one asteroid mass PBH inspiral per year, $\tilde N(d)>1$, the maximal distance at which the binaries are observable must exceed
\be \label{eq:dmin}
	d \approx 0.1 \,{\rm kpc} \left[\frac{L}{\rm m}\right]^{\frac23} \left[\frac{A}{{\rm m}^2}\right]^{-\frac13} \left[\frac{m}{10^{20}\,{\rm g}}\right]^{\frac{32}{111}} \,.
\ee
If the closest PBH binary is at that distance, the production rate of induced photons is
\bea
	\Gamma_\gamma(t) \approx & \frac{0.002}{\rm day}\!\left[\frac{\bar B_x}{10\,{\rm T}}\right]^2 \!\left[\frac{L}{\rm km}\right]^{\!\frac23} \!\left[\frac{A}{100{\rm m}^2}\right]^{\!\frac53} \\ &\times\!\left[\frac{m}{10^{20}\,{\rm g}}\right]^{\!\frac{121}{111}} \!\left[\frac{f}{{\rm GHz}}\right]^{\!\frac23} ,
\eea
where we have normalized the detector characteristics in a way that the rate is non-negligible. We conclude that detecting GWs from PBH binaries via graviton-to-photon conversion is beyond the reach of near future experimental capabilities or requires an accidental nearby source.

A similar study~\cite{Herman:2020wao} estimated that the signal is many orders of magnitude stronger than ours. The discrepancy stems from the assumed experimental concept design. We assume that the measurement is on the induced photon flux, while Ref. ~\cite{Herman:2020wao} proposes to measure the variation in the energy of the cavity due to the cross term between the external and induced magnetic fields (Eq.~(XIV) of Ref.~\cite{Herman:2020wao}). As a result, they find that the signal is linear in the GW strain, while the photon flux \eqref{eq:Ps} is quadratic. They also include a resonant enhancement of the signal, but that gives a subdominant contribution to the discrepancy. To the extent that the concept design of Ref.~\cite{Herman:2020wao} is realizable, it would certainly represent a very significant improvement in sensitivity.

\section{Conclusions}

We have studied the prospects of probing PBHs with future GW experiments. We have shown that planned GW interferometers can probe a wide mass range from $10^{-6}M_\odot$ to $10^9M_\odot$ and, moreover, provide the most sensitive probe of all currently available PBH observables in the slightly narrower mass range $10^{-5}M_\odot$ to $10^7M_\odot$. In particular, we found that observing the inspiral phase of PBH binaries in the Milky Way can be used to probe subsolar mass PBHs with masses as low as $10^{-5}M_\odot$ and abundances as low as $f_{\rm PBH} = 10^{-3}$. Furthermore, searches for galactic BH binaries by ET and AEDGE can independently test the PBH explanation for the six ultrashort microlensing events observed by OGLE. At the heavier end of the mass spectrum, the SGWB observations can be used to study heavy $10^{3}M_\odot-10^{7}M_\odot$ PBH binaries at abundances lower than $f_{\rm PBH} = 10^{-5}$.

We also considered the intriguing possibility of observing high-frequency GWs from asteroid mass PBH binaries through graviton-photon conversion and found that such PBHs are likely out of reach of near-future experiments.

\vspace{2mm}
\emph{Acknowledgments:} 
This work was supported by the Spanish MINECO grants FPA2017-88915-P and SEV-2016-0588, the grant 2017-SGR-1069 from the Generalitat de Catalunya, the European Regional Development Fund through the CoE program grant TK133, the Mobilitas Pluss grants MOBTP135, MOBTT5, and the Estonian Research Council grant PRG803. IFAE is partially funded by the CERCA program of the Generalitat de Catalunya.

\bibliography{PBH}
\end{document}